\documentclass[twocolumn,showpacs,preprintnumbers,amsmath,amssymb]{revtex4}
\usepackage{graphicx}% Include figure files
\usepackage{dcolumn}% Align table columns on decimal point
\usepackage{bm}% bold math
\usepackage[english]{babel}

\newcommand{\nphysa}{Nucl. Phys.~A }%
\begin{document}

\title{
Spherical configuration of a super--dense hot compact object with particular EoS
}
%% Running heads
%\shorttitle{Origin of heavy elements} \shortauthors{Tito, Pavlov}

%\author{E. P. Tito}
%\affil{Scientific Advisory Group, Pasadena, California, USA}
%\author{V. I. Pavlov}
%\affil{UFR des Math\'{e}matiques Pures et Appliqu\'{e}es-- LML
%CNRS UMR 8107, \\ Universit\'{e} de Lille 1, 59655
%Villeneuve d'Ascq, France \\
%\vspace{0.4cm}
%\textnormal{November 16, 2013}}

\author{E. P. Tito}
%\email{eptito@gmail.com}
\affiliation{Scientific Advisory Group, Pasadena, CA 91125, USA}

\author{V. I. Pavlov}
\affiliation{Univ.~Lille,~UFR~des~Math\'ematiques~Pures~et~Appliqu\'ees,~CNRS~FRE~3723~-~LML,~F-59000~Lille,~France}

\date{October 08, 2015}

\begin{abstract}
The equation of state (EoS) $P = P (\rho, ...)$  -- pressure as a function of density and other thermodynamical quantities  -- is what generates particularities of mass--radius distribution $M (R)$ for super--dense compact stellar bodies, the remnants of cosmic cataclysms. In view of recent nuclear experiments, we propose one particular EoS, which admits the critical state characterized by density $\rho_c$ and temperature $T_c$, and which under certain conditions permits
a radial distribution of the super--dense matter in "liquid" phase. We establish such conditions and demonstrate that a stable configuration is indeed possible (only) for temperatures smaller than the critical one. Using  Tolman--Oppenheimer--Volkoff equations for hydrostatic equilibrium, we derive the mass--radius relation for the super--dense compact objects with masses smaller than the Sun, $M \ll M_{\odot}$. The obtained results are within the constraints established by both heavy--ion collision experiments and theoretical studies of neutron--rich matter.
    % "La folie est de faire et de refaire la m\^eme chose en esp\'erant des r\'esultats diff\'erents" (Albert Einstein).
\end{abstract}

%\pacs{47.20.Cq, 47.10.Df, 47.20.Ma}
%\pacs{98.80.Bp, 47.10.-g, 95.30.Lz}

\maketitle

\section{Introduction} \label{S:Introduction}

The  equilibrium spherical configuration of the non--rotating self--gravitating system is found from the set of Tolman--Oppenheimer--Volkoff (TOV) equations \cite{ov39, t39},
well known to be:
\begin{eqnarray}
\frac{d P}{d r} = - \gamma_N \frac{m \rho}{r^2} ( 1 + \frac{P}{ c^2
\rho})(1 + 4 \pi \frac{P r^3}{m c^2}) ( 1 - 2 \frac{\gamma_N m}{c^2 r}
)^{-1}, \label{sdco01b} \\
\frac{d m}{d r} = 4 \pi r^2 \rho .  \label{sdco01a}
\end{eqnarray}
Here, all quantities are in usual units, $P = P(r)$ is the pressure at radius $r$, $\rho$ is the mass density of the matter which includes all forms of energy together with the rest mass, $\gamma_N$ is the newtonian gravitational constant, $c$ is the light speed.  Quantity $m$ in Eq.(\ref{sdco01a}) is the "mass inside radius $r$":
\begin{eqnarray}
m  = 4 \pi \int_0^r d s s^2 \rho (s) .  \label{sdco02}
\end{eqnarray}
The total mass of the body, $M$, is the integral of Eq.(\ref{sdco02}) from $0$ to $R$.\footnote{
This integral includes all contributions to the mass including gravitational potential energy \cite{st83}: in fact, the proper volume element in the gravity field is not  $4 \pi  r^2 d r$ but $4 \pi  r^2 g_{rr}^{1/2} d r$, i.e. $4 \pi  r^2 (1 - 2 m / r)^{-1/2} d r$. When $r \rightarrow R$, $m$ must become equal to $M$, so that the interior metric matched smoothly the exterior Schwarzschild metric.
}
Terms $\geq c^{-2}$ and $\geq \gamma_N^2 $ after decomposition in series with respect to $c^{-1}$ and $\gamma_N $ of right part of
Eq.~(\ref{sdco01b}), give contributions produced by the effects of the special and general theories of
relativity.\footnote{
The gravity potential (the parameter connected with the space--time metrics) is found from
${d \Phi}/{d r} = -({1}/{\rho}) ({d p}/{d r}) ( 1 + {p}/{c^2 \rho} )^{-1} . \label{sdco:02c}$
}

To obtain the density distribution and the mass--radius relationship $M (R)$ for both ordinary stars and dense compact objects, for a chosen equation of state (EoS) $P(\rho, \, ...)$, Eqs.(\ref{sdco01b}) -- (\ref{sdco02}) must be integrated subject to boundary conditions $P(R)=0$ and $m (0)=0$. Obviously, the dependence $M (R)$ is strongly model--dependent on the form of the EoS -- a crucial point in this problem.

\begin{figure}[ht!]
\centering
\includegraphics[width=7cm]{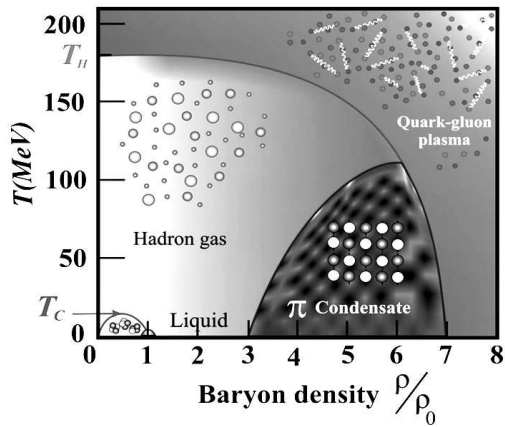}
\caption{The phase diagram for nuclear matter, as predicted theoretically \citep{g84}.
%(A. L. Goodman et al., Phys. Rev.C 30 (1984) 851).
The horizontal axis shows the density of the matter. The vertical axis shows its temperature. Both axes are given in logarithmic scale. The density is given in multiples of normal nuclear matter density  $\rho_0 \simeq 2.8 \times 10^{17} \, kg/m^3$ (neutron drip). For densities higher than $\rho_0$, the nuclei begin to dissolved and merge together by forming the neutron liquid. In the range of higher densities,  physical properties of the matter are uncertain.  At present, it is believed that hadronic matter, at high enough densities, undergoes a transition to a deconfined state of quarks and gluons
\citep{w84}, \citep{fj84}.
%(Witten 1984; Farhi \& Jaffe 1984).
In a general context, it is assumed that as the density--temperature increase above the "normal" nuclear characteristics, the matter may undergo phase transitions to qualitatively new states (meson condensation, crystallization, quark deconfinement, mixed phases, etc.): the very existence of these states depends on the specific features of strong interactions and the quark structure of baryons.
\label{phasediagram}}
\end{figure}

\begin{figure}[ht!]
\centering
\includegraphics[width=7.0cm]{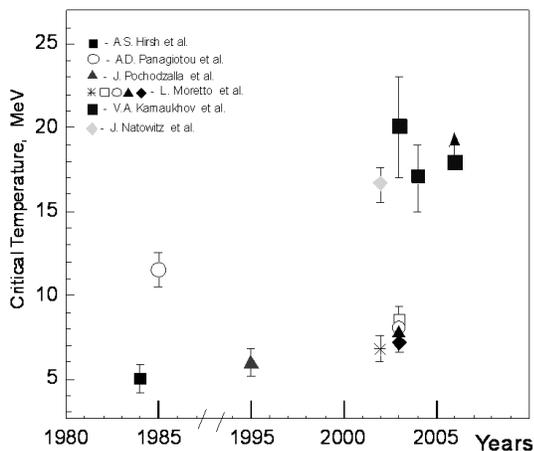}
\caption{
Values of critical temperatures of nuclei $T_c$ measured by different techniques. From
\citep{k-a11}.
}
 \label{TcKarn11}
\end{figure}

In this paper, for the EoS in our consideration, we additionally allow for two possibilities: the possibility that $P (\rho_k)=0$ for some $\rho_k \neq 0$, and
the possibility of the existence of the critical state corresponding to some density $\rho_c$ and some temperature $T_c$. The realization of the latter has been supported by nuclear experiments on collisions of heavy nuclei (see Figs.~\ref{phasediagram} and  \ref{TcKarn11}).

In this paper, we establish conditions in which a  super--dense matter (governed by such EoS) can exist in its "liquid" phase in a form of stable radial distribution (like a droplet). We show also that such stable configuration is possible only for temperatures lower than the critical one.

By the term "super-dense compact object" (SDCO), we  call a gravitationally-powerful stellar body, a remnant of a cosmic cataclysm, whose average density is of order of the nuclear density~\footnote{
"Normal" nuclear density is $\rho_0 \sim 2.8 \times 10^{17} \, kg/m^3$; the density in the center of a SDCO must be an order of magnitude higher than $\rho_0$
},
whose mass is (meaningfully) smaller than the Sun mass, and whose physical state  in the inner region can be modeled by an EoS permitting the multi--phase state (the region in Fig.~\ref{phasediagram} where $T < 20 \, Mev$ and $\rho < 3 \div 4 \, \rho_0$).

The physics of the compact star population (white/black dwarfs, traditional neutron stars, hyperon stars, strange stars and possible quark stars or so-called hybrid stars) involves a complicated interplay between nuclear processes and astrophysical phenomena
(\cite{z61},
\citep{hf75},
\citep{st83},
\citep{s89},
\citep{cls90},
\citep{ls91},
\citep{u95},
\citep{w95},
\citep{g97},
\cite{apr98},
\citep{stos98b},
\citep{dh01},
\citep{k06},
\citep{hpy07},
\citep{f09},
\citep{p10},
\citep{slb12},
\citep{fr12},
\citep{tp13}).
The  equation of state (EoS) -- the dependence of  pressure $P$ on  energy (mass) density $\rho$ -- is central to the calculation of compact star's properties as it determines the mass range, the  mass--radius relationship, and other characteristics \cite{slb12}.\footnote{For example, measuring of gravitational red--shift $Z = (1- 2 \gamma_N M_b (R_b) / R_b c^2 )^{-1/2} - 1$ permits  measuring, in principle, radius $R_b$ and allows to verify the acceptability of a particular model of EoS. Here, $M_b$ is the mass of body, $R_b$ is its radius, $\gamma_N$ is the gravitational constant, $c$ is the light speed.}

The rationale for the possible form of EOS has been based on data obtained from diverse sources, such as studies of high energy nuclear collisions, the monopole resonance in finite heavy nuclei, astrophysical supernovae and neutron star studies \cite{wgw91}, \cite{prl95}, \cite{lrp93}. The general concept of the contemporary understanding is illustrated in Fig.~\ref{phasediagram}. The critical temperature $T_c$ and the critical density of energy $\varepsilon_c$ which is proportional to critical density $\rho_c$, can be measured  nowadays in nuclear experiments. Fig.~\ref{TcKarn11} presents the various experimental data.

Obviously, of importance for SDCO  are the effects of the general relativity theory (GR). The significance of the effects of gravity for a body of radius $R_b$
with the inner parameter $\varepsilon_c$ (critical energy density), is determined by the dimensionless parameter $G = \gamma_N R_b^2 \varepsilon_c / c^4$ (see below) which is small for physically--interesting situations. We take into account that the high--density matter of any stable object must conform with
(i) causality (the speed of propagation of small matter perturbations must not exceed the speed of light),
(ii) hydrodynamical stability.

In Sec.~2, we describe the model which permits a radially--symmetrical distribution of mass (the choice of an equation of state, the dimensionless formulation of the set of equations, the necessary thermodynamical quantities). The configuration of non--rotating stars is described by the relativistic equation of hydrostatic equilibrium for a spherically symmetric body - the TOV--equations \cite{ov39, t39}. In  Sec.~3, we find the matter density distribution. The causality condition (the speed of sound must not exceed the speed of light) is discussed in Sec.~4. Mass--Radius dependence for the SDCO is obtained in Sec.~5.  A special case is briefly considered in Sec.~6. In the concluding Sec.~7, we discuss the obtained results.

\section{Radially-symmetrical distribution of mass}

\subsection{Equation of state}

The basic equations are written above, Eqs.~(\ref{sdco01b})--(\ref{sdco02}). The principal point for the following analysis is to propose a physically reasonable explicit form of the equation of state (EoS). In fact, the exact EoS of dense matter remains a well--kept secret of nature in spite of decades of very intense theoretical and experimental studies (see more detailed discussion for example the works by \citet{hpy07, f09, p10}.

A reasonable model for the nuclear matter EOS must be thermodynamically self--consistent and reproduce such quantities as normal concentration $n_0$ and  incompressibility factor $K$ in the vicinity of normal nuclear matter. At normal nuclear density the free energy must be at minimum (then the system is mechanically stable because of vanishing pressure $P = P_0 = 0$) and the width of this minimum is defined by the incompressibility factor $K$. The behavior of the model in other regions of the $n-T$ plane can be then probed via heavy--ion collisions.

An important point is the so--called ground state of nuclear matter, i.e., at $T = 0$ nuclear matter saturates (pressure $P =P_0 = 0$) at a concentration  of about $n_0 \simeq 0.16 \, fm^{-3}$. The nucleon-nucleon interaction is generally attractive at nucleon--nucleon separations of $> 1 \, fm \, (= 10^{-15} m)$ but becomes repulsive at small separations ($< 0.5  \, fm$) making nuclear matter difficult to compress. As a consequence, most stable nuclei  are at approximately the same saturation density, $\rho_0 \simeq 2.8 \times 10^{14} \, g / cm^3$, in their interiors. Matter at densities of up to
$\rho \simeq 9 \rho_0$ may be present in the interiors of neutron stars, and matter at densities up to about $\rho \simeq 4 \rho_0$ may be present in the core collapse of Type~II supernovae.

In \citep{j83} (see also \citep{g84} and \citep{s04}), the  empirical nuclear equation of state (EOS), polynomial approximation,
\begin{equation}
\label{n01}
P = \frac{T}{m} \rho - A_1 \rho^2 + A_2 \rho^3 ,
%p = T n - b n^2 + c n^3 ,
\end{equation}
has been proposed to explain observed experimental data. Here, $T$  is temperature,
%is measures in energetic units,
$\rho = \varepsilon / c^2$, $\varepsilon$ is energy density of the matter,
%(in classic limit) is written as $\rho = m n$, $n$ is concentration of particles,
$A_1 = T_c / m \rho_c$ and $A_2 = 2 T_c / 6 m \rho_c^2$.
%$b = T_c / n_c$ and $c = 2 T_c / 6n_c^2$.
Such  EoS permits the existence of the critical state when the first and second derivatives of pressure with respect to density turn zero (surface tension vanishes). Coefficients $b$ and $c$ depend directly on the value of the critical temperature $T_c$, critical density $\rho_c$. Despite the fact that a heavy nucleus is a "repository" of strongly interacting fermions, and the first term in this formula refers to the classical system, Eq.~(\ref{n01}) works well and permits to satisfactorily describe   results of experiments. This is explained by the fact that the finite system of strongly interacting fermions is satisfactorily described in terms of Green's function as an ensemble of (localized in finite volume) collective perturbations (of the spin zero) which have the non--zero effective "mass" and the degree of excitation of which can be characterized by some effective "temperature".

The meaning of Eq.~(\ref{n01}), or other interpolating expressions below, is as follows: In the gas state under normal conditions when $p \sim \rho$, interaction between particles is very weak. As the interaction (pressure) increases, the properties of the system differ more and more from the properties of the ideal gas, and finally the gas enters its condensed state -- liquid. In the liquid state, interaction between particles is great, and properties of this interaction strongly depend on the specific type of the liquid. This is the reason why general formulae, describing quantitatively properties of liquids, do not exist (see \cite{ll69}--\cite{ll59}). However, it is possible to propose some {\em interpolation} formula which can qualitatively describe the transition between the gas and liquid (as done in the Van der Waals classical model). Such formula must produce qualitatively correct results in two limit cases. For rarified gases it should converge into formulae correct for ideal gases. But as the density increases, it should incorporate the fact that the compressibility of the matter is limited. Such formula then would qualitatively describe the gas behavior in the transition state.

Eq.~(\ref{n01}) represents only one of the numerous possible interpolation formulae satisfying the posed requirements. There are no physical reasons to prefer one such interpolation over the others. But the form (Eq.~\ref{n01}) is one of the simplest and easiest to work with.

\begin{figure}[ht!]
\centering
\includegraphics[width=6.5cm]{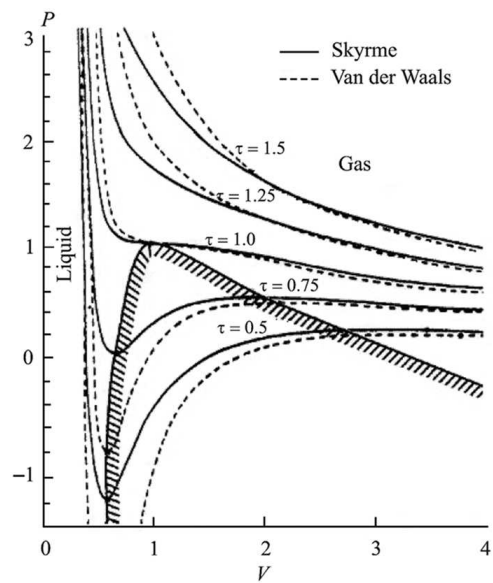}
\caption{
The equations of state $P(V)$ for a nuclear system interacting through a Skyrme potential and a Van der Waals compressible liquid--gas system  (shown in relative units). (From~\citep{j83}.)}
\label{nucl-phase2}
\end{figure}

\begin{figure}[ht!]
\centering
\includegraphics[width=7cm]{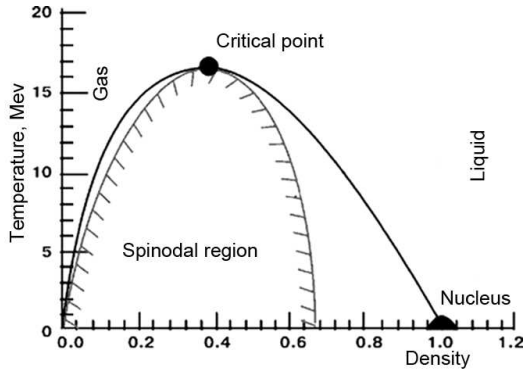}
\caption{Theoretical $T(\rho)$ phase diagram for nuclear matter (adapted from \citep{k06}). The solid line is determined by condition $\partial p / \partial \rho = 0$ and marks the phase transition zone. Density is expressed in units of  $\rho_{nucleus} \simeq 2.85 \times 10^{14} \, g/cm^3 $. Temperature is expressed in $Mev$ units ($1 \, Mev \simeq 10^{10} K$).}
\label{spinodal-reg}
\end{figure}

The equations of state of a multi--body system of nucleons interacting via Skyrme potential is presented in Fig.~\ref{nucl-phase2}.
The very steep part of the isotherms (on the left side) corresponds to the liquid phase. The gas phase is presented by the right parts of the isotherms where pressure is changing smoothly with increasing volume. Of special interest is the part of the diagram where the isotherms correspond to the negative compressibility, i.e. $(\partial P / \partial V )_T > 0$. This is the so-called {\em spinodal} zone where the matter phase is unstable and can exist in both liquid and/or gas states.\footnote{Following to Karnaukhov V. A. et all
%"Properties of Hot Nuclei Produced in Relativistic Collisions" (17.10.2011)
\citep{k-a11}: "Spinodal decomposition $=$
liquid--fog phase transition".}
Within the spinodal zone lies  a particularly unstable two--phased region (marked by the hatched line in Fig.~\ref{spinodal-reg}), in which random density fluctuations lead to almost instantaneous collapse of the initially uniform system into a mixture of two phases. For nuclear matter, it is either liquid droplets surrounded by gas of neutrons, or homogeneous neutron liquid with neutron--gas bubbles (i.e. the spinodal zone where the squire of adiabatical speed is negative, is inside the coexistence zone where the squire of isothermical speed is negative).
%WHICH ZONE IS INSIDE WHICH ???????

Critical temperature $T_c$ for the liquid-gas phase transition is a crucial characteristic of the nuclear equation of state.

\begin{figure}[ht!]
\centering
\includegraphics[width=8.5cm]{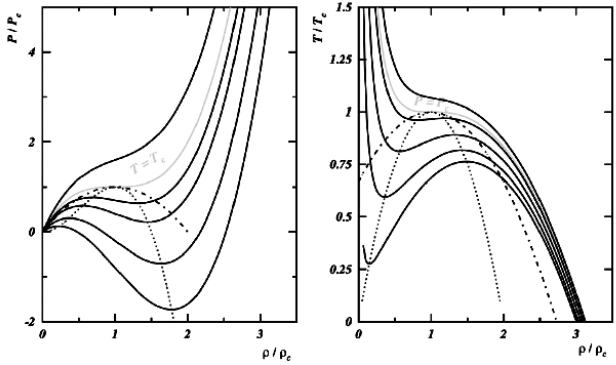}
\caption{
Equation of state for nuclear matter:  pressure (isotherms, left panel) or  temperature (isobars, right panel) as functions of  density.
(Parameters are normalized by their critical values).
The dash-dotted lines are the coexistence lines, the dotted lines are the spinodal lines.
From
\citep{b02}.
See also, \citep{br08}.}
\label{br08}
\end{figure}

A typical set of isotherms for an equation of state (EoS) - pressure versus density with a constant temperature - corresponding to nuclear interaction  (Skyrme effective interaction and finite temperature of Hartree--Fock theory, see
\citep{j83})
is shown in Fig.~\ref{br08}.
It exhibits the maximum-minimum structure typical of the VdW--like EoS. Depending on the effective interaction chosen and on the model
(see \citep{j83}, \citep{j84}, \citep{c86}, \citep{m95}),
the nuclear equation of state exhibits a critical point at $\rho_c \simeq (0.3 \div 0.4)\rho_0$ and $T_c \sim 5 \div 18 \, MeV$
(\cite{k06}, \cite{k-a11}).
Calculations of $T_c$ were performed in
\cite{g84}, \cite{j83}, \cite{s04},  \cite{s76}, \cite{z96}, \cite{t04}.
Experimental data are presented in Fig.~\ref{TcKarn11}.

Some model EoS -- pressure vs. internal energy density -- are given in Appendix~\ref{ap:1}.

\subsection{Dimensionless TOV equations}

We introduce dimensionless variables $r \rightarrow R_b s , \; \rho \rightarrow \rho_c z (s), \; \varepsilon = \varepsilon_c \, \varepsilon (s), \; \rho_c = \varepsilon_c /c^2 ,\; m \rightarrow ( \varepsilon_c R_b^3 / c^2 ) m (s)$.
Then the dimensionless quantity  $m (s)$  is defined as
\begin{eqnarray}
m (s) = 4 \pi \int_0^s d \xi \xi^2 z (\xi) , \label{eos:10a}
\end{eqnarray}
and $M_b$ becomes
\begin{eqnarray}
M_b = 4 \pi \rho_c R_b^3 \int_0^1 d s s^2 z (s). \label{eos:10b}
\end{eqnarray}

We suppose that the pressure is measured in units $n_c T_c$, i.e. $P \rightarrow n_c T_c p(s)$.
Then  Eqs.~(\ref{sdco01b}) and (\ref{sdco01a}) for dimensionless  quantities $m$, $z$ and $p$ become
\begin{eqnarray}
\frac{d m}{d s} - 4 \pi s^2 z = 0,  \label{eos:13a}\\
\frac{d p}{d s} +
\frac{1}{T} \frac{G}{s^2} ( z  + T p  ) ( m  + 4 \pi T s^3 p )
%\nonumber\\ \times
( 1 - 2 G \frac{m}{s} )^{-1} = 0.
\label{eos:13b}
\end{eqnarray}
Here, the dimensionless parameters $T = T_c / m_n c^2 \simeq 0.0186$ when $T_c = 17.5 \, Mev$, $m_n = 939.76 \, Mev$ and $G = \gamma_N \rho_c R_b / c^2 \simeq 0.01024 a^2 / z_k$ with $R_b = 10^{-6} a$.
The system of obtained equations with an EoS $p = p (\theta , z)$ contains now only one dimensionless parameter $G  \sim a^2$ which is defined by the size of the SDCO, $a$.

If $G \ll 1$, we write in Eqs.~(\ref{eos:13a}) and (\ref{eos:13b}) $( 1 - G f )^{-1} \simeq ( 1 + G f + ...)$ and neglect further small terms of order $G^4$ (and higher). The newtonian approximation corresponds to contribution of terms $\sim G^1$.
The effects of relativity are taken into consideration in such approximated version of Eqs.~(\ref{eos:13a}) and (\ref{eos:13b}) -- terms with $p$
in right part of Eq.~(\ref{eos:13b}) following from special relativity, and terms $\simeq G^2$ following from the GR.
In  the definition of the small parameter $G$, $\gamma_N$ is the gravity constant, $c$ is the light speed,
dimensionless $m = m_n / T_c$, $\varepsilon_c = \rho_c T_c / m_n$.

The set of equations (\ref{eos:13a}) -- (\ref{eos:13b}) is subject to boundary conditions (see below).

\subsection{Thermodynamical quantities}

All principal thermodynamical quantities such as pressure, internal energy, sound speed and so on, can be calculated when a thermodynamical potential is given (see Appendix, Eqs.(\ref{mod04}). Sometime, this is the free energy of the system which is preferable for description of a specific system. To illustrate,
we consider the simplest mono--component system composed of $N$ particles occupying volume $V$. Following the definition, the full free energy of a system is a function of temperature $T$ and of volume, $F = F(T, V, N)$. With respect to one particle, with finite $n = N / V$ even when $N \rightarrow \infty, \; V \rightarrow \infty$ (thermodynamical system), the fundamental thermodynamical relationship is written as $d F_1 = - s_1 d T - P d (n^{-1})$ with the chemical potential $\mu_1$ per one particle defined from $F_1 + P n^{-1} - \mu_1 = 0$.

Introducing the dimensionless thermodynamical arguments $z = n / n_c$, $\theta = T / T_c$, $F_1 = T_c f$, $P = n_c T_c p$, we have the expressions
\begin{eqnarray}
p = z^2 \partial_z f, \, s_1 = - \partial_{\theta} f, \, u_1 = - \theta^2 \partial_{\theta} (f / \theta).   \label{eos:tqa}
\end{eqnarray}
The dimensionless expression for internal energy per volume unit is calculated as
\begin{eqnarray}
\varepsilon = z u_1 = - z \theta^2 \partial_{\theta} (f / \theta).   \label{eos:tqb}
\end{eqnarray}

We propose the following simple expression for the dimensionless free energy $f = F_1 / T_c$ per one particle
\begin{eqnarray}
f =  m  - \theta \ln \bigg( \frac{\theta^{3/2}}{z} \bigg)  +
a_1 \, Li_2  (- a_2 z) +  \frac{a_3}{2} z + \frac{3}{2} \theta .   \label{eos:fe}
\end{eqnarray}
Here, $m, a_1, a_2, a_3$ are dimensionless constants, the function $Li_2 (-x)$ is the polylogarithm function of argument $x$. In limit cases, $Li_2 (-x) \simeq - x + x^2 / 4 - x^3 / 9 + \, ...$ for $x \rightarrow 0$, and $Li_2 (-x) \simeq \pi^2 / 6  - (1/2) \ln^2 (1 / x) + (1 / x) \, ...$ for $x \rightarrow \infty$. The physical meaning of every term in $f$ and numerical values of constants $a_1, a_2, a_3$ will be discussed later.
The last term is added to set the value of entropy equal to zero at the critical point. This additional term changes nothing in measurable quantities (pressure, internal energy, etc).

The model Eq.(\ref{eos:fe}) must produce the following results:
(a) the EoS following from Eq.(\ref{eos:fe}) has to have a form permitting the existence of the critical point where $p = \partial_z p = 0$;
(b)  pressure  $p (z_1) = 0$ when $z_1 \neq 0$;
(c) the critical density $\rho_c$ is of order of $(0.1 \div 0.4) \, \rho_0$, i.e. $z_1 \simeq (3 \div 7)$;
(d) compressibility factor $K \sim (240 \div 300) \, Mev$;
(e) the principle of causality is respected - the adiabatical sound speed is always smaller than the light speed - $V_s^2 < 1$.

The dimensionless pressure and volume density of internal energy are calculated now as (see Appendix \ref{ap:3})
\begin{eqnarray}
p = z \theta - a_1 z \ln (1 + a_2 z) + \frac{a_3}{2} z^2,  \label{eos:06}\\
\varepsilon = m z + \frac{3}{2} z \theta + a_1 z \, Li_2 (- a_2 z) + \frac{a_3}{2} z^2,  \label{eos:07}
\end{eqnarray}
Here, pressure and internal energy per volume unit contain terms which are proportional to effective temperature as for the classical ideal gas. Simple consideration shows that the free energy per one particle $f$ cannot be a polynome  of power higher then $1$. In fact, let, for large $z$, the leading term in $f$ be proportional to some power of $z$, i.e. $f \sim z^{\nu-1}$ with $\nu >1$. Since (see Appendix~\ref{ap:3} and Eqs.~(\ref{eos:06}) and (\ref{eos:07})),
$p = z^2 \partial_z f \sim (\nu-1) z^{\nu}$ and $\varepsilon = - z \theta^2 \partial_{\theta} (f / \theta) \sim z^\nu$ with the same coefficients of proportionality, the  adiabatical sound speed (normalized by the light speed square) $V^2_s = (\partial p / \partial \varepsilon)_s \rightarrow (\nu-1)$ for $z \rightarrow \infty$. This gives the one possibility: $\nu = 2$. Obviously, it follows from here that the interpolating function in Eq.~(\ref{eos:fe}) must tend to zero, when $z \rightarrow 0$, not so fast as a linear function of $z$, and tend to infinity not faster then $z^1$ when $z \rightarrow \infty$.

Expression (\ref{eos:07}), as well as  Eq.~(\ref{eos:06}), has a simple physical meaning: the first term $m z$ is determined by the rest mass of particles, the second is connected with heat motion, the third describes interaction (attraction) of particles for moderate density of matter, the last term is connected with repulsion due to hard "core" inside particles. If there are no "wonders" in the region of moderate densities, $z \sim 1$, the interpolating function can be taken in the simple (quasi-logarithmic) form. The physical reasons of such comportment are the subject of more detailed and specific investigation which are outside the framework of this paper.

The constants $a_1, a_2, a_3$ are fixed by the following three conditions: (a, b) the first and second derivatives of the pressure $p (z, \theta)$ are zero in the critical point $z = 1, \, \theta = 1$; (c) the dimensionless factor of incompressibility must give the experimentally obtained value.

\begin{figure}[ht!]
\centering
\includegraphics[width=0.8\columnwidth]{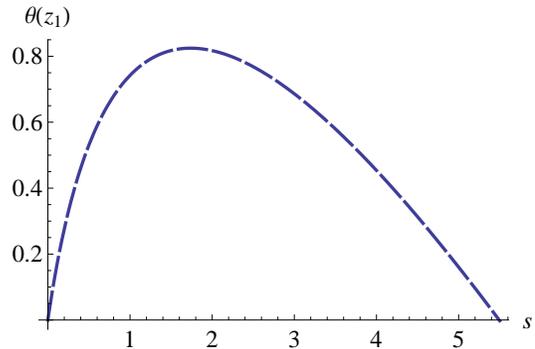}
\caption{Density $z (1, \theta)$ at the boundary $s = 1$ as a function of temperature $\theta$.
The spherical configuration for the liquid phase exists when $0 < \theta < \theta_{max} = 0.826$ and when $z_{min} (1) = 1.74 < z(1) < z_{max} (1) = 5.5$ (the right part of the curve).
\label{temperature-size}}
\end{figure}

The value of $z(1)$ for different temperatures $\theta$ is found from the condition at the boundary $p (z(1), \theta) = 0$. The physical solution of this equation exists not for all values of parameters. The dependence $z (1, \theta))$, or $\theta = \theta (z(1))$, is shown in Fig.~\ref{temperature-size}.

The important parameter for any EoS is the incompressibility factor $K$. It measures the stiffness of the EOS, usually defined as a slope of the pressure at the point of fundamental state (saturation point): $K \sim ( \partial_n P )_{T \rightarrow 0, \, n \rightarrow n_0}$ (in usual units).
This expression (in its dimensionless form used in the paper) is written as $\kappa = (9 z^2 \partial_{z z} \varepsilon )_{\theta \rightarrow 0, \, z \rightarrow z_0}$. In fact, $\kappa = 9 (\partial_z p)$ is equal to $\kappa = 9 \partial_z (z^2 \partial_z f) = 9 z^2 \partial_{zz} f$ in the fundamental state (when $\theta = 0$ and $z = z_1$ where $p (z_1, 0) = 0$). When $\theta \rightarrow 0$, we can use $f = \varepsilon$.  The module of inelasticity is expressed in $Mev$ when $T_c$ is given in $Mev$: $K = T_c \kappa$.  The experimental value of the parameter is of order $200 \div 300 \, Mev$
 (see for example
 \cite{g88})
 with a value of $K = 300 \, MeV$ (with considerable error), or $K = 180 \div 240 \, MeV$  from
 \cite{b80},
 \cite{tkbm81}.
 Some experimental value of the incompressibility of symmetric nuclear matter at its saturation density $n_0$ has been determined to be $210 \pm 30 \, MeV$
\cite{hpy07}.
So, the situation is still not very clear: the analysis of the sideward anisotropy observed in the heavy ions collisions at low and intermediate energies require $K \simeq 210 \, MeV$, whereas the the elliptic flow anisotropy observed in the same experiment requires $K \simeq 300 \, MeV$
%[62, 63]
\cite{dll02},
\cite{d05}.

In the framework of our model when the critical temperature is taken as $T_c = 17.5 \, Mev$, we find for parameters $a_1, a_2, a_3, z_1$ numerical values $a_1 = 1.225, a_2 = 1.841, a_3 = 1.074, \, z_1 = 5.5$ with the factor of incompressibility $\kappa= 16.533$ (i.e. $K = 289.3 \, Mev$),  which is a satisfactory result given the approximations made along the way.

\begin{figure}[ht!]
\centering
\includegraphics[width=0.8\columnwidth]{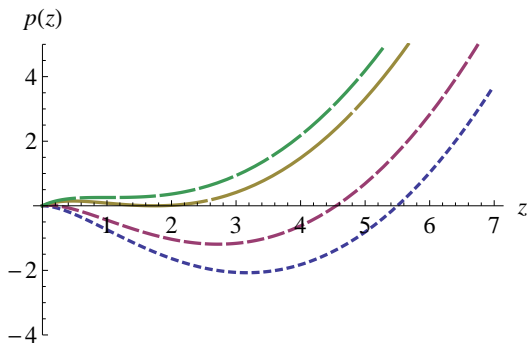}
\caption{
Pressure $p (z, \theta)$ as a function of normalized density $z$ for several values of normalized temperature $\theta$:
 $\theta =0$ (lowest line), $\theta= 0.3$ (second line from bottom), $\theta = 0.8255$ (second line from top) with point where $p = \partial_z p = 0$, and critical isotherm $\theta = 1.0$ (upper line).
All curves below the critical isotherm, i.e. when $0 < \theta < 1$, possess two turning points ($z_1 < z_2$) where  $(\partial_z p )_{z=z_i} = 0$.
In the domain $0 < z < z_1$, the matter is in its gas state. In the domain $z > z_2$,  the matter is in its liquid state.
Between $z_1$ and $z_2$, lies the zone, where the gas and liquid phases co-exist.
}
\label{Press}
\end{figure}

The behavior of  pressure $p (z, \theta)$ calculated from Eqs.(\ref{eos:tqa}) and (\ref{eos:fe}), is shown in Fig.\ref{Press} for different temperatures.

\begin{figure}[ht!]
\centering
\includegraphics[width=0.8\columnwidth]{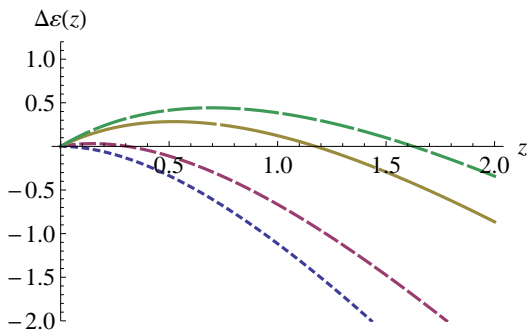}
\caption{
Quantity $\Delta \varepsilon (z, \theta) = \varepsilon (z, \theta) - m z$ as a function of normalized density $z$ for several values of normalized temperature $\theta$:
 $\theta =0$ (lowest line), $\theta= 0.3$ (second line from bottom), $\theta = 0.8255$ (second line from top) with point where $p = \partial_z p = 0$, and the curve corresponding to critical temperature $\theta = 1.0$ (upper line).
}
\label{EnIntFig2A}
\end{figure}

\begin{figure}[ht!]
\centering
\includegraphics[width=0.8\columnwidth]{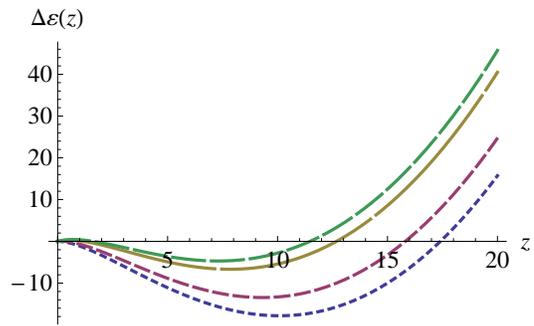}
\caption{
Quantity $\Delta \varepsilon (z, \theta) = \varepsilon (z, \theta) - m z$ as a function of normalized density $z$ for large densities and  for several values of normalized temperature $\theta$:
 $\theta =0$ (lowest line), $\theta= 0.3$ (second line from bottom), $\theta = 0.8255$ (second line from top) with point where $p = \partial_z p = 0$, and the curve corresponding to critical temperature $\theta = 1.0$ (upper line).
}
\label{EnIntFig2B}
\end{figure}

The behavior of  internal energy  per volume unit is shown in Figs.\ref{EnIntFig2A} and \ref{EnIntFig2B}.

\begin{figure}[ht!]
\centering
\includegraphics[width=0.8\columnwidth]{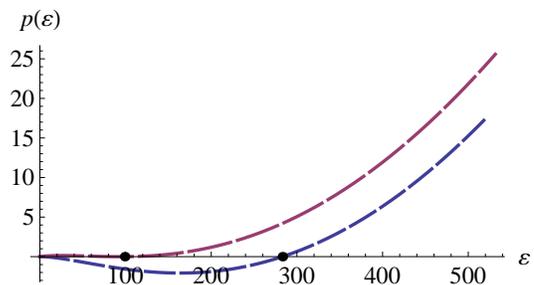}
\caption{
Pressure versus internal energy density of the matter.
 \label{pe-relationFig}}
\end{figure}

\begin{figure}[ht!]
\centering
\includegraphics[width=0.8\columnwidth]{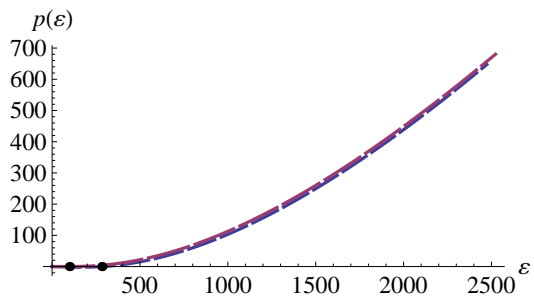}
\caption{
Pressure versus internal energy density of the matter for large values of $z$. The factor of transformation to usual units is $T_c n_c = 0.525 \, Mev /fm^3$.
 \label{pe-relationGFig}}
\end{figure}

Fig.\ref{pe-relationFig} shows the pressure $p$ vs internal energy density $\varepsilon$ relation for the matter in liquid phase
with limit temperatures $\theta_1 = 0.8255$ and $\theta_0 = 0$ for which the liquid state, $\theta_0 < \theta < \theta_1$, can exist.
When $z \gg 1$, the limit curves tend to the universal relation $p = \varepsilon$.
The proposed model is not in contradiction with the existing models (see Fig.\ref{Fig-wnr07}). For example, according to our model, when $e = T_c n_c \varepsilon = 1050 \, Mev / fm^3$ (with $T_c = 17.5 \, Mev$ and $n_c = 0.03 \, fm^{-3}$), we obtain for the pressure $P = T_c n_c p = 210 \, Mev / fm^3$, i.e. the $p (\varepsilon)$ dependence shown in Fig.\ref{pe-relationGFig} is near the RMF~(npKH)~model.

\section{Speed of sound}

The adiabatical sound speed (i.e. speed of propagation of small "acoustical" perturbations $=$ the speed of transport of energy $=$ the speed of transport of information) for very high densities must not be greater than the light speed.

The adiabatical (dimensionless here) sound speed  for relativistic fluid is calculated using the expression $V^2_s =( \partial p / \partial \varepsilon )_s$.
This quantity is calculated in condition that the entropy per  one particle $s$ is constant. However, the pressure and internal energy in the model are functions of density $z$ and temperature $\theta$.
Therefore, it is more natural to calculate $V_s^2$ using the Jacobians  and their properties (see \citep{ll69}, \citep{rr77} for more information).
Using the Jacobians we can find the expression

\begin{eqnarray}
\bigg( \frac{\partial p}{\partial \varepsilon} \bigg)_s \equiv \frac{\partial (p, s)}{\partial (\varepsilon, s)}
=
\frac{ p_z  - s_z  ( s_{\theta} )^{-1} p_{\theta} }{ \varepsilon_z  - s_z  ( s_{\theta} )^{-1} \varepsilon_{\theta}}
.
\label{mod08b}
\end{eqnarray}

\begin{figure}[ht!]
\centering
\includegraphics[width=0.8\columnwidth]{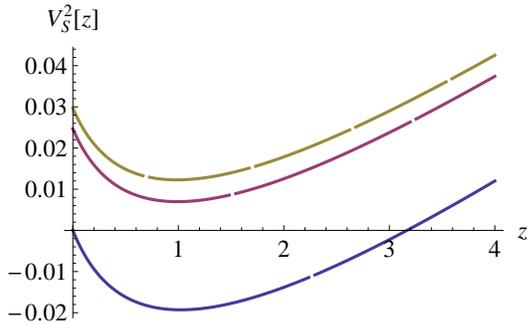}
\caption{
Square of adiabatical sound speed $V^2_s (z)$, normalized by the speed of light, as a function of normalized density $z$ for several values of normalized temperature $\theta$: $\theta = 1$ (upper line), $\theta =0.8355$ (the curve $p (z, \theta_*)$ touches the horizontal axis in plane ($z, p$)) and $\theta =0$ (lower line).
The domain with $V^2_s (z) < 0$, where the sound speed $V_s(z)$ is imaginary, is the so-called "spinodal" zone. The condition $V^2_s (z) < 0$
indicates that small spontaneous initial perturbations of density will grow exponentially fast.
The development of the instability in a homogeneous medium leads to formation of a two--phase configuration when liquid (drops) and gas (vapor) states co--exist.  Only the states corresponding to temperatures below some temperature $\theta_*$ (unique for the medium),
for which the curve $V^2_s (z)$ touches the horizontal axis in plane ($z, V^2_s$),
possess such domain of instability.
For the states with $\theta > \theta_*$, the speed of sound is always real ($V^2_s (z) > 0$) and the matter exists in a mono--phase state.
}
\label{V2b}
\end{figure}

\begin{figure}[ht!]
\centering
\includegraphics[width=0.8\columnwidth]{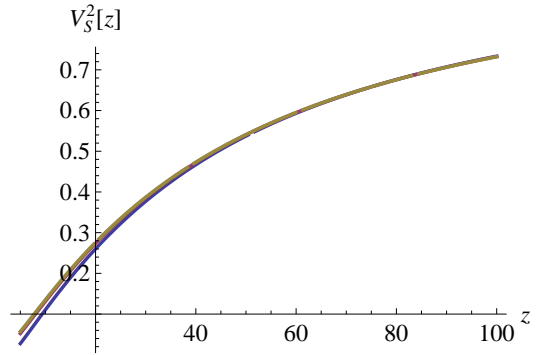}
\caption{
"Zoom-Out" for Fig.~\ref{V2b}. Square of adiabatical sound speed $V^2_s$ as function of (large) $z$ for $\theta = 1$ (upper line), $\theta =0.8355$ and $\theta =0$ (lower line). In the proper framework, the speed of sound cannot exceed the speed of light: $V^2_s \leq 1$. Bethe-Johnson or Van der Waals equations of state violate this requirement. In our model, the sound speed correctly tends (at large z) to the speed of light (unit of one).
}
\label{V2c}
\end{figure}

Eqs.(\ref{mod04}), (\ref{mod08b}) with (\ref{eos:fe}) permit calculating of all derivatives in Eq.~(\ref{mod08b}) and  finding $V_s^2$. Dimensionless speed of sound in our model always satisfies  the condition $V^2_s < 1$ (Fig.~\ref{V2c}). The system becomes unstable with respect to small spontaneous perturbations (fluctuations) when $V^2_s < 0$ (Fig.~\ref{V2b}).

\begin{figure}[ht!]
\centering
\includegraphics[width=0.8\columnwidth]{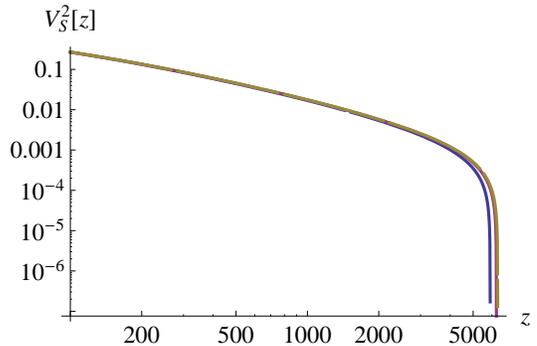}
\caption{
"Zoom-Out" for Fig.~\ref{V2c}.
The quantity $1 - V^2_s$ as function of (large) $z$ in logarithmic scale.
As pictured, in our model, the sound speed correctly tends (at large z) to the speed of light (unit of one) and does not exceed it.
}
\label{V2d}
\end{figure}

\begin{figure}[ht!]
\centering
\includegraphics[width=0.8\columnwidth]{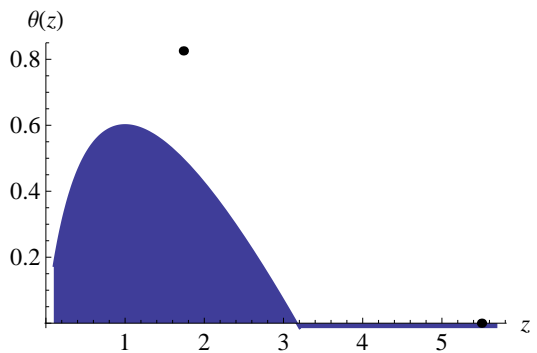}
\caption{
Spinodal region in plane ($z, \theta$) (inside the domain, $V_s^2 < 0$; outside the domain, $V_s^2 > 0$). }
 \label{Spinod}
\end{figure}

\begin{figure}[ht!]
\centering
\includegraphics[width=0.8\columnwidth]{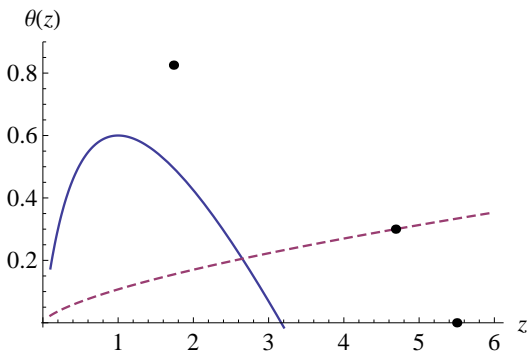}
\caption{
Spinodal region in plane ($z, \theta$) (inside the domain, $V_s^2 < 0$; outside the domain, $V_s^2 > 0$). The coordinates of some points in which the pressure is zero are shown: $(5.5, 0); (4.687, 0.3); (1.74, 0.8255)$. Any process pushing the system from initial state  ($z_0, \theta_0$) into the spinodal region adiabatically (line $\theta = \theta_0 (z / z_0)^{2/3}$), leads to the development of collective instability and to  the fragmentation of matter.}
 \label{SpinReg}
\end{figure}

\section{Radial density distribution}
Eqs.~(\ref{eos:13a}), (\ref{eos:13b}), (\ref{eos:06}) and (\ref{eos:07}) complete the system of equations from which a radially--symmetrical distribution of mass within the SDCO can be found. The set of the equations is subject to the boundary conditions $p(1)=0, m(0)=0$. At this point, there is a difference with traditional approaches when Eqs.~(\ref{eos:13a}) -- (\ref{eos:13b}) are numerically integrated for a given central density.

We write in Eqs.~(\ref{eos:13a}),(\ref{eos:13b}), (\ref{eos:06}) and (\ref{eos:07})
$z  = z (1) + G z_1 (s) + G^2 z_2 (s) +  G^3 z_3 (s) + ...$ and $m = ({4 \pi}/{3})z(1) s^3 (1 + G \mu_1 (s) + G^2 \mu_2 (s)  + G^3 \mu_3 (s)+ ... )$. Here, $z(1)$ is the normalized on $\rho_c$ density at $s =1$ (i.e. on the boundary where pressure is assumed zero; $p(z(1))= 0$). Quantities $z_i$ and $\mu_i$ may be
considered as "add-ons", small perturbations of the basic state.

The rough estimation of the validity of such consideration can be written as
\begin{eqnarray}
2 G \frac{m (s)}{s} \leq \frac{8 \pi}{3} G z(0) \simeq 0.07 a^2  ,
\end{eqnarray}
for the hypothetically  taken $z(0) \sim 10^1 z(1)$, i.e. for $a \ll 3 \div 4$.

The first boundary condition is obviously $z_i (1) = 0$. The second is not $\mu_i (0) = 0$ because $z(1) \neq z(0)$: when $s \rightarrow 0$, $m (s)$ has to tend to ${4 \pi}/{3})z(0) s^3 + ...$ in leading approximation. The boundary condition for $\mu_i (s)$ when $s=1$, is chosen after solving the set of equations from condition that the quantity $m(s)$ has no singularities at any point in the region $0 \leq s \leq 1$. Thus, $\mu_i (1)$ are the proper values of the set equations.

After substitution of these expressions into Eq.~(\ref{eos:13b}), and equating coefficients with the same power of $G^n$, we obtain the set of equations for the normalized density add-on $z_i (r)$ and mass add-on $\mu_i (r)$ which is resolved by the standard procedure.

\section{Mass--radius relationship}

\begin{figure}[ht!]
\centering
\includegraphics[width=0.8\columnwidth]{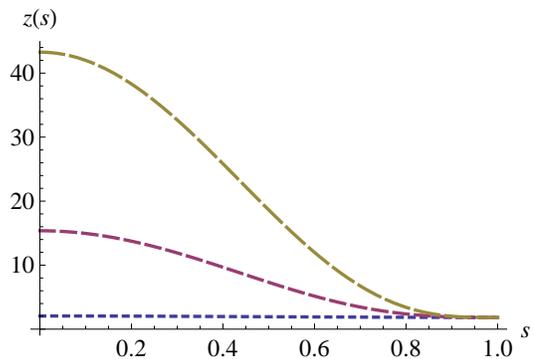}
\caption{
Density distribution for a SDCO (with temperature $\theta =0.824$) as a function of distance from center $s$ for different values of the SDCO radii $a$: $a = 0.05$ (lower line), $a = 0.1$ and $a = 0.12$ (upper line).
}
\label{zs-config}
\end{figure}

\begin{figure}[ht!]
\centering
\includegraphics[width=0.8\columnwidth]{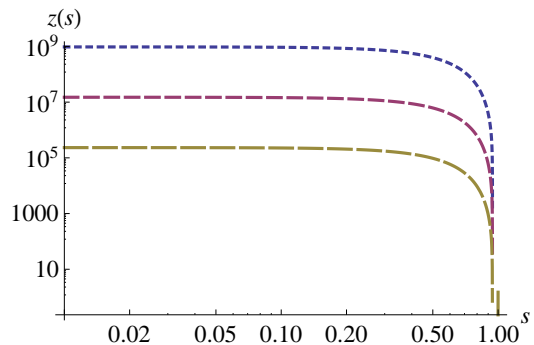}
\caption{
Density distribution for a SDCO (with temperature $\theta =0.824$) as a function of distance from center $s$ for large values of $a$: $a = 0.5$ (lower line), $a = 1$ and $a = 2$ (upper line). For such great density, the density distribution is quasi-homogeneous and is essentially defined by the GR effect.
}
\label{LDH}
\end{figure}

The density distribution for the hot SDCO with $\theta =0.824$, is shown in Figs.(\ref{zs-config}) and (\ref{LDH}).

The mass--radius relationship follows from Eq.~(\ref{eos:10a}).  To simplify the final expressions, we introduce special units: the mass $M_b$ of a SDCO will be measured in $10^{-3} M_{\odot}$ units and the radius, $R_b = 10^{-6} R_{\odot} a$, is $a$ units. Here, the Sun's mass and radii are denoted by $\odot$. Numerically, $R_\odot = 7 \times 10^8 \, m$  the mass of the Sun $M_\odot \simeq 2 \times 10^{30}\, kg$, $\gamma_N = 6.67 \times 10^{-11} \, $, $c = 2.99 \times 10^7 \, m / s$,  $\rho_c \simeq 0.35 \rho_0$.

The mass--radii relationship becomes
\begin{eqnarray}
\frac{M}{10^{-3} M_{\odot}} = 0.211 a^3 \int_0^1 d s \, s^2 \, z(s; a, \theta) .
\end{eqnarray}
It can be also presented in the form
\begin{eqnarray}
\frac{M}{10^{-3} M_{\odot}} = 0.168 a^3 m(1; a, \theta) \nonumber \\
\equiv 0.168 \frac{4 \pi}{3} z[1] a^3 F(a, \theta, z[1]).
\end{eqnarray}

\begin{figure}[ht!]
\centering
\includegraphics[width=0.8\columnwidth]{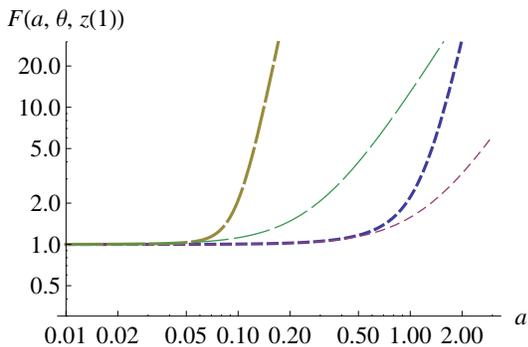}
\caption{
Parameter $F (a, \theta, z[1])$ defining the mass--radius relationship  (in $10^{-3} M_\odot$ units for mass and $10^{-6} R_\odot$ for size) for cold ($\theta = 0$) and hot ($\theta = 0.824$) SDCO.
The domain where $F \simeq 1$ corresponds to the case when one can neglect the GR gravity effect. The thin dotted lines correspond to the newtonian approximation.
    %???????  NEED TO DISTINGUISH THE LINES, and EXPLAIN BETTER ??????????????
}
\label{factF}
\end{figure}

\begin{figure}[ht!]
\centering
\includegraphics[width=0.8\columnwidth]{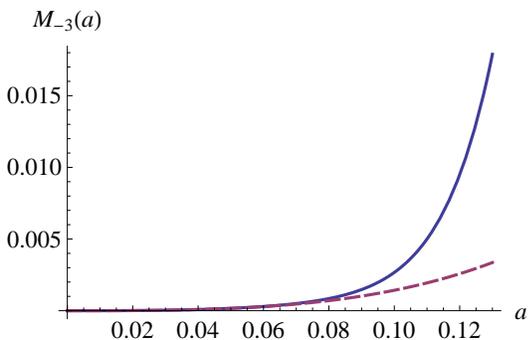}
\caption{
Mass--radius relationship  (in $10^{-3} M_\odot$ units for mass and $10^{-6} R_\odot$ for size) for a hot SDCO (with temperature $\theta =0.824$). Dashed line plots the relationship in the newtonian approximation.
}
\label{mass-rad}
\end{figure}

Mass--radius relationship $M_{-3} (a)$ (mass in $10^{-3} M_\odot$ units) for the considered special case when the temperature $\theta=0.824$, i.e. near the critical temperature $T_c \sim 17.5 \, Mev$, is shown in Fig.~\ref{mass-rad}.

\section{Conclusion}

We considered conditions for which SDCOs -- small stellar bodies with a specific nuclear equation of state -- can exist in the nature.
The SDCO's mass as a function of its size,  has been obtained.

The principal point of the work is the interpolating expression for the dimensionless free energy (Eq.~\ref{eos:fe}) from which all thermodynamical quantities can be found. The expression for the free energy reflects the following limit conditions: for small densities, $z \rightarrow 0$,  the interaction between particles is weak, and the dominant term is the first term which describes a gas of non-interacting particles. As the density increases, the properties of the system differ more and more from the properties of the ideal gas, the interaction (logarithmic term in expression for pressure)  becomes more and more significant. With more increasing of density, $z \gg 1$, the gas enters its condensed state (liquid) when the term $\sim z$ in expression for $f$ is the most important. For high densities $z$, the equation of state has to be "hardened" to account for the dominance of the "repulsive core" in the potential of particle interaction. In such "hardened" state, repulsion between particles is very strong, and the properties of this interaction no longer depend on the specific type of the liquid, thus the corresponding term in the free energy has to have a universal form for the pressure $p \sim z^2$ corresponding to arguments of \citep{z61}.
%Zel'dovich (1961)

The proposed model Eq.(\ref{eos:fe}) responded to the following requirements:  %exigences:
(a) the EoS following from Eq.(\ref{eos:fe}) has to have a form admitting the existence of the critical point where $p = \partial_z p = 0$; (b) the pressure  $p (z_1) = 0$ for some value $z_1 \neq 0$;
(c) the critical density $\rho_c$ is of order of $(0.1 \div 0.4) \, \rho_0$, i.e. $z_1 \simeq (3 \div 7)$;
(d) compressibility factor $K \sim (240 \div 300) \, Mev$;
(e) the principle of causality is respected - the adiabatical sound speed is always smaller than the light speed.

The proposed model of the EoS permitted to construct a spherical self--gravitating configuration: the SDCO.

In Figs.~\ref{Press} - \ref{mass-rad} we graphically illustrate some of the obtained results: the equations of state $p(z,\theta)$ and the square of sound speed $V^2_s (z, \theta)$ for various values of temperature $\theta$ (and scale zooms for $z$), resulting from the model (Eq.~\ref{eos:fe}). The calculations have been made with parameters $a_1 = 1.225, a_2 = 1.841, a_3 = 1.074$. The figures demonstrate the existence of the spinodal zone for temperatures below critical, where the square of the sound speed is negative. This signifies that the speed of sound is imaginary in the domain, indicating that small spontaneous initial perturbations of matter density grow exponentially fast in beginning of process. The instability process leads to formation of the liquid--gas phase state. The model correctly captures the principle of causality when the speed of propagation of small perturbation of matter density is smaller then the light speed.

Any process pushing the system from initial "liquid" state  ($z_0, \theta_0$) into the spinodal region for example adiabatically (following to lines $\theta = \theta_0 (z / z_0)^{2/3}$), leads to instability development and fragmentation. To obtain such a situation, it is sufficient to "rarify" some domain of the SDCO  (for example $\rho_0 \rightarrow \rho_0 / 2$). It can be accomplished  by a sharp deceleration of the SDCO.

A stationary spherical configuration exists only if the boundary condition for pressure $p = 0$ is respected for some $z_1 \neq 0$. With respect to
Fig.~\ref{Press}, this signifies that there is intersection of curves $p = p (z, \theta_1)$ for given $\theta_1$ with horizontal axis $p = 0$. The $z_1 \neq 0$ is the boundary value of density which corresponds to $p (z_1 , \theta_1 ) = 0$. If some mechanism (for example due to a simple deceleration of the object colliding with another massive object) introduces some quantity of heat into the SDCO, the system passes into a new state characterized by a new value of temperature, $\theta_1 \rightarrow \theta_2 > \theta_1$. In this case, the curve of $p (z, \theta_2)$ may not intersect the horizontal axis $p = 0$: instead, for example, of the middle or lowest lines in Fig.~\ref{Press}, the configuration will be characterized by the upper line. This signifies that an equilibrium spherical configuration for the SDCO does not exist more: all is manifested as an explosion of the system and its destruction in multitude of fragments which can be unstable too with respect to specific nuclear reactions.

The model admits a generalization to a multi--component system, including thermic radiation. For this, the expression for the free energy must be replaced by the sum of the free energy expression for every component.

\appendix

\section{Actually existing models of the EoS \label{ap:1}}

\begin{figure}[ht!]
\centering
\includegraphics[width=\columnwidth]{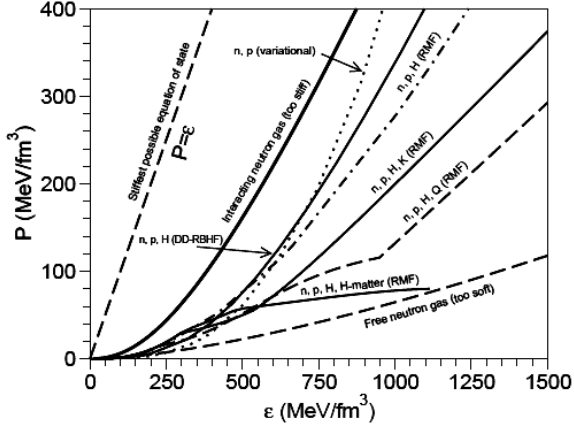}
\caption{
Some models for pressure versus energy density of neutron
matter (from \citep{wnr07}). The notation is as follows: RMF is the relativistic mean-field
model; DD-RBHF is the density dependent relativistic Brueckner-Hartree-Fock model;
"n" are neutrons, "p" are protons, "H" are hyperons, "K" is is the $K [u,s]$ meson condensate; "Q" signifies $(u, d, s)$
quarks; "H-matter" is the H-dibaryon condensate.
 \label{Fig-wnr07}}
\end{figure}

A number of models for the EoS of neutron matter have been presented in literature over the years.
According to \citep{wnr07} where the references are presented, these models can roughly be classified as follows:
Thomas-Fermi based models; %[41, 42],
Schroedinger-based models (e.g. variational approach, Monte Carlo techniques,
hole line expansion (Brueckner theory), coupled cluster method,
Green function method); % [8, 43, 44, 45, 46, 47, 48]
Relativistic field-theoretical treatments (relativistic mean field (RMF),
Hartree-Fock (RHF), standard Brueckner-Hartree-Fock (RBHF), density
dependent RBHF (DD-RBHF); % [49, 50, 51, 52, 53, 54]
Nambu-Jona-Lasinio (NJL) models; % [55, 56, 57, 58, 59, 60]
Chiral SU(3) quark mean field model . %[61].
A collection of equations of state computed for several of these models is shown
in Fig.~ \ref{Fig-wnr07}.

All presented here models show a monotone dependence of pressure as function of energy density in the region of moderate values, and the two--phase state of matter is not realized in framework of the models.

\section{Thermodynamical potentials and related quantities}
\label{ap:3}

\subsection{Thermodynamical observables}

Different thermodynamical quantities can be used in the macroscopical description of the system.

The fundamental thermodynamic relationship (FTR) for the change of the internal energy $U$ of a system is  $d U = T dS - P dV + \mu d N$. Here, all quantities have the  standard meaning and dimension, $T$ is the temperature of the system, $S$ is the entropy, $P$ is pressure, $V$ is the full volume, $\mu$ is the chemical potential, and $N$ is full number of particles. It follows from here that  $U = U (S, V, N)$.

For the free energy $F = U - T S$, we obtain from the first law of thermodynamics that $d F = - S dT  - P dV + \mu d N$. From here it follows that the free energy is a function of temperature, volume, and number of particles: $F = F (T, V, N)$. If we introduce quantities reported to one particle, $F = F_1 N$, $S = s_1 N$, $V = M / m_n n$, $N = n V$ where $M$ is the full mass of the system, $V$ is its volume, $m_n$ is the nucleon mass,  we can write that the free energy reported to one particle is $F_1  = - P / n + \mu_1 $ and its differential is $d F_1 = - s_1 d T - P d (1 / n)$. The pressure is thus a function of temperature and particle concentration: $P (T, n) = n^2 (\partial F_1 / \partial n)$. The chemical potential and the internal energy also per one particle are $\mu_1 = \partial_n (n F_1)$ and $u_1 = - T^2 \partial_T (F_1 / T)$. The internal energy reported to volume unit (which has the same dimension as pressure) is given by $E_1 = n u_1 = - n T^2 \partial_T (F_1 / T)$.

For the \emph{dimensionless} quantities defined by expressions $F_1 = T_c f , \, n = n_c z, \, P = T_c n_c p , \, T = N_c \theta , \, \mu_1 = T_c \mu, \, u_1 = T_c u, \, E_1 = T_c n_c \varepsilon$, we can write
\begin{eqnarray}
\label{mod02}
d f = - s d \theta - p d ({1}/{z})
\end{eqnarray}
Pressure $p$, entropy $s$ and the volume density of internal energy $\varepsilon$ are calculated as
\begin{eqnarray}
\label{mod04}
p = z^2 \partial_z f , \quad  s = - \partial_{\theta} f, \quad \varepsilon =  - z \theta^2 \partial_{\theta} \frac{f}{\theta} . \quad
\end{eqnarray}
Note, if the free energy $f$ is given as a function of density $z$ and temperature $\theta$, the other thermodynamical quantities depend on $z$ and $\theta$ too.

In some situations, the so-called grand canonical potential, $\Omega$, also called the Landau free energy, or Landau potential, is more useful. The quantity $\Omega$ is defined via the full free energy $F$ of the system as $\Omega = F - \mu N = U - T S - \mu N$. The change of the grand potential is found from definition of $\Omega$ and the FTR� and is given by $d \Omega = - S d T - P d V - N d \mu $. This expression shows that the grand canonical potential is a function of parameters $V, T, \mu$: $\Omega = \Omega (V, T, \mu)$. Letting $N = n \, V$, $S = \overline{s} \, V$ and $\Omega = \omega \, V$, we obtain that $V \, (d \omega + \overline{s} \, dT + n d \mu)= - (\omega + P) d V$. It follows from here that, for homogeneous thermodynamical systems, $P = - \omega (T, \mu)$, or $\Omega = - P V $  and $d \omega = - \overline{s} \, d T - n d \mu$ \cite{ll96}. The density of particles is given by derivative $n (T, \mu) =  (\partial P / \partial \mu )$, the entropy reported to the unit of volume is $\overline{s} (T, \mu) = (\partial P / \partial T)$ and the internal energy (in terms $T, \mu$) reported to volume unit is $\overline{u} (T, \mu) \equiv (U/V) = - P  + T (\partial P / \partial T) + \mu (\partial P / \partial \mu)$. The combination $\overline{u} (T, \mu) - T \overline{s} (T, \mu)$ gives a volume density of generalized free energy $\overline{f} = \overline{f} (T, \mu)$ with $\overline{f} = - P + \mu (\partial P / \partial \mu)$ which is equally expressed in terms $T, \mu$: the function $\overline{f}$ has the same dimension as pressure $P$. Resolving $n = n (T, \mu) \rightarrow \mu = \mu (T, n)$, one finds $\overline{f}$ in traditional form $\overline{f} (T, \mu (T, n))$

\end{document}